\documentclass[12pt]{article}
\usepackage{amsmath}
\usepackage{graphicx}
\usepackage{amsfonts}
\usepackage{amssymb}
\begin{document}

\title{Localization and superconducting proximity effect in sandwiched potassium films}
\author{Manjiang Zhang and Gerd Bergmann\\University of Southern California}
\date{\today}
\maketitle
\begin{abstract}
Thin films of alkali metals when sandwiched at both surfaces by thin metal
films loose their conductance. The superconducting proximity effect is used to
investigate the change in the alkali film. On the length scale of the film
thickness the electronic properties of the alkali film do not change
noticeably although its conductance is dramatically reduced, corresponding to
localized electrons.

PACS: 74.45.+c, 73.20.Fz, 71.20.Dg

\newpage
\end{abstract}

In recent years we have studied thin films of alkali metals with spin-orbit
impurities, magnetic impurities and (s,p)-impurities and observed a number of
surprizing phenomena \cite{B111}, \cite{B121}, \cite{B122}, \cite{B123},
\cite{B124}, \cite{B129}, \cite{B130}, \cite{B131}, \cite{B132}. Among the
different results there was one finding that was particularly puzzling. When
we covered a Cs film with sub-mono-layers of impurities the resistance of the
film increased dramatically \cite{B107}. We tried many different impurities
such as Pb, In, Au, Ag, etc. They produced essentially the same result. In
Fig.1 the effect of In surface impurities on the conductance of a Cs film is
shown. In addition to the resistance the Hall constant increased as if the
density of conduction electrons decreased or the thickness of the alkali film
was reduced.

$%
\raisebox{--0.0166in}{\parbox[b]{3.7634in}{\begin{center}
\fbox{\includegraphics[
natheight=3.113300in,
natwidth=4.143600in,
height=2.8352in,
width=3.7634in
]%
{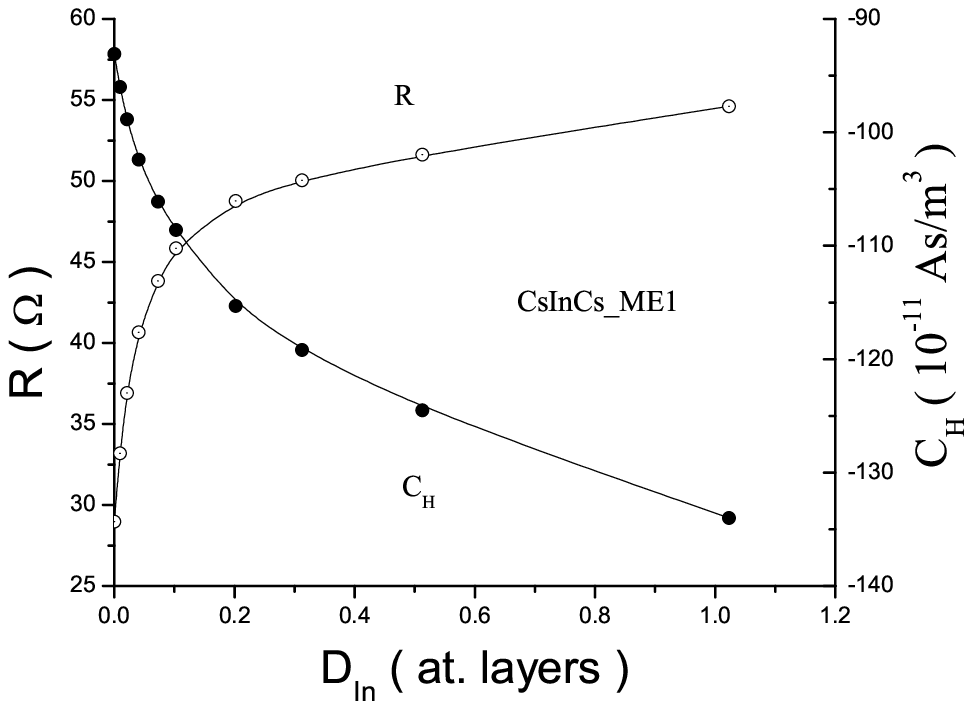}%
}\\
Fig.1: The resistance and the Hall-constant of a Cs film, covered with
sub-mono-layers of In, as a function of the In coverage.
\end{center}}}%
$

There is a related difference between film of alkali and other metals: If one
condenses at He temperatures an insulating film of amorphous Sb or MgF$_{2}$
and covers such a film with a non-alkali metal, for example with Pb, then this
film becomes already conducting for film thicknesses between 1 and 4 atomic
layers. On the other hand if one condenses an alkali metal, for example Cs on
top amorphous MgF$_{2}$ the first 15 atomic layers are non-conducting (while
on quartz one can observe conductance for a thickness of 5 layers.)

This observation raises the question whether it is possible to transform an
alkali film into an non-conducting state by covering it with impurities.
Indeed we observed such an effect in thin layers of Cs and K covered with Sb
or Pb. We first condensed a 3.2 $nm$ thick insulating film of MgF$_{2}$ (at
helium temperatures) onto a quartz plate. On top of the fresh substrate a K
film of 6.5 $nm$ (about 18 atomic layers) and a resistance of 1082 $\Omega$
was condensed. Then the K was covered with sub-mono-layers of Pb. In table I
the resistance is given as a function of the Pb coverage. For a Pb coverage
between 1.0 and 1.5 atomic layers the resistance diverges.%

\[%
\begin{tabular}
[c]{ll}%
\begin{tabular}
[c]{|l|l|}\hline
\textbf{d}$_{\text{Pb}}$ (a.l.) & \textbf{R (}$\Omega)$\\\hline
0.0 & 1082.\\\hline
0.018 & 1234.\\\hline
0.048 & 1549.\\\hline
0.1 & 2110.\\\hline
0.21 & 3200.\\\hline
0.51 & 5621.\\\hline
1.0 & 25860.\\\hline
1.51 & $\infty$\\\hline
\end{tabular}
&
\begin{tabular}
[c]{l}%
$\text{Table I: The resistance }$\\
$\text{of a K film as a function }$\\
$\text{of the coverage with Pb.}$%
\end{tabular}
\end{tabular}
\]

Our quench condensed alkali films are very clean, possessing mean free paths
up to 1000A (five time the thickness of the films) since all the evaporation
sources are surrounded with liquid N$_{2}$ and that the vacuum in our system
is better than 10$^{-11}$torr. Nevertheless we looked for supportive experiments.

We believe we have found an experiment which more clearly demonstrates the
unusual behavior of the alkali films. It is a combination of transport
properties together with the superconducting proximity effect. We prepare a
sandwich by the following sequence of evaporations: (a) an insulating film of
10 atomic layers of amorphous Sb, (b) on top of the amorphous Sb a Pb film
with a thickness of 6$nm$ and a resistance of 33.\thinspace$\Omega$ is
condensed (the freshly condensed Sb substrate makes it possible to obtain a
homogeneous flat film of Pb, avoiding the formation of islands). This thin
layer of Pb has a transition temperature of 6.88$K$. (c) Then the Pb is
covered in three steps with K which is a normal conductor. As one expects for
a superconductor-normal conductor (SN) sandwich the transition temperature
decreases with increasing K thickness and reaches the value of 4.79$K$ for a K
thickness of 4.9$nm$. This is shown in Fig.2. (d) Finally the K is covered
with Pb in steps of .02, .05, .01, .3, 1.0, .. mono-layers. Initially the Pb
has no effect on the transition temperature. Only when the Pb thickness
reaches one mono-layer the $T_{c}$ increases slightly. The likely reason for
this increase in $T_{c}$ is that one approaches an SNS sandwich.

$%
{\parbox[b]{3.907in}{\begin{center}
\fbox{\includegraphics[
natheight=3.058500in,
natwidth=3.823200in,
height=3.1299in,
width=3.907in
]%
{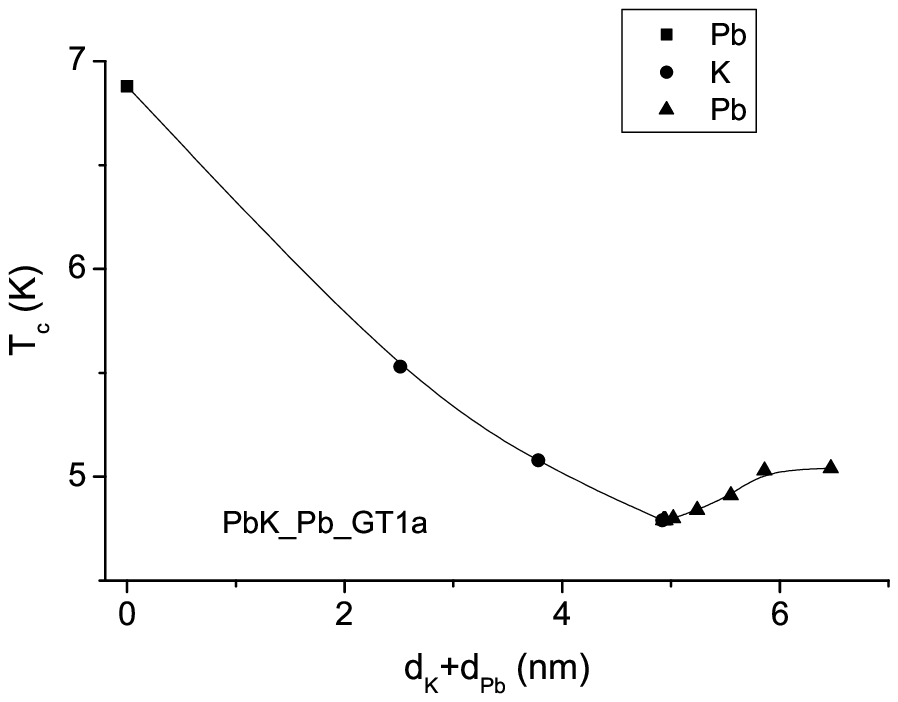}%
}\\
Fig.2: The superconducting transition temperature of a sandwich consisting of
(Sb)/Pb/K/Pb.
\end{center}}}%
$

The observed behavior in $T_{c}$ is typical for an SN sandwich and agrees
qualitatively with the transition temperature of PbCu sandwiches \cite{H25},
\cite{H26}, \cite{M45}. What is very unusual is the behavior of the resistance
(conductance) of the sandwich. This is shown in Fig.3 where the conductance is
plotted versus the coverage with K and (the second) Pb. The evaporation of the
first K film decreases the conductance. That means that (i) the K increases
the resistance of the Pb film and (ii) contributes nothing (or very little)
itself to the conductance. This changes for the second and third K coverage.
One observes a clear increase in the conductance due to the K film.

$%
{\parbox[b]{3.8464in}{\begin{center}
\includegraphics[
natheight=2.978000in,
natwidth=3.837300in,
height=2.9913in,
width=3.8464in
]%
{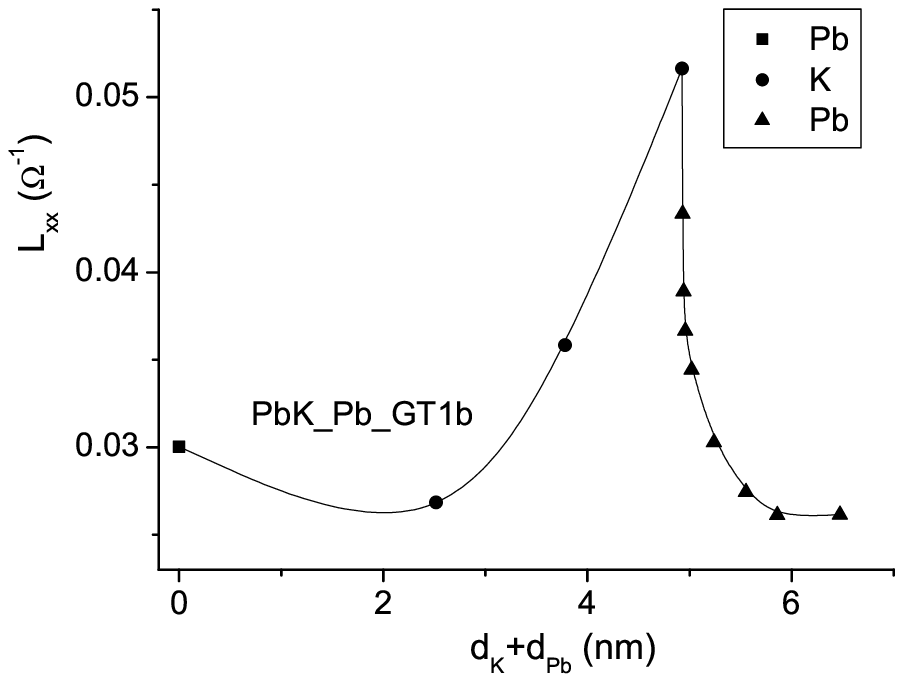}%
\\
Fig.3: The conductance of a the same sandwich as in Fig.2 as a function of the
K and Pb coverage.
\end{center}}}%
$

The evaporation of the sub-mono-layers of Pb reduces the conductance of the
sandwich. With increasing Pb coverage the conductance of the sandwich drops
not only below the conductance of the pure Pb film but even below the minimum
value of the Pb/K sandwich. The conductance of the sandwich behaves as if the
whole K film becomes insulating. At the same time the Hall conductance of th
sandwich decreases. This is shown in Fig.4.

$%
{\parbox[b]{3.7791in}{\begin{center}
\includegraphics[
natheight=3.005400in,
natwidth=4.037300in,
height=2.8202in,
width=3.7791in
]%
{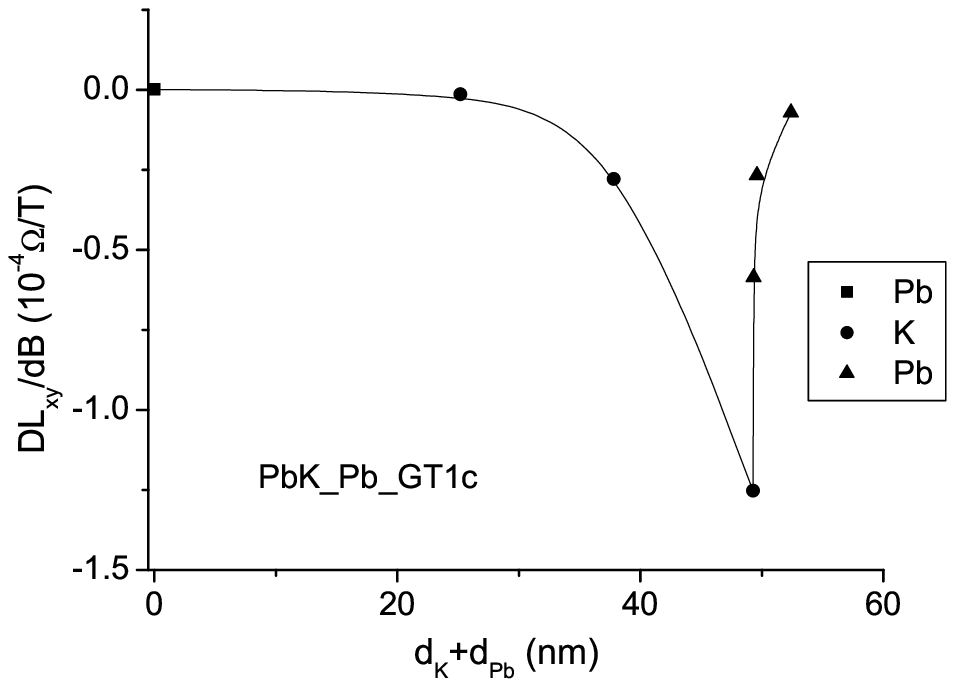}%
\\
Fig.4: The linear Hall conductance $dL_xy/dB$ of a the same sandwich as in
Fig.2 as a function of the K and Pb coverage.
\end{center}}}%
$

An advocator diablo might say that together with the Pb some O$_{2}$ is
introduced into the system, oxidizing the K so that the effective thickness of
the K is reduced. We can exclude this possibility because it would result in
an increase of the transition temperature of the sandwich: the effective
thickness of the normal conductor would be reduced and therefore the $T_{c}$
would approach the $T_{c}$ of the first Pb film. However, the $T_{c}$ of the
sandwich remains practically constant.

In the following we want to show that the unaffected transition temperature of
the sandwich excludes that

\begin{itemize}
\item  the effective thickness of the K film is reduced

\item  the electron density in the K film is reduced

\item  the mean free path perpendicular to the film is noticeable reduced.
\end{itemize}

The properties of an inhomogeneous superconductor are determined by the gap
equation \cite{G38}. Close to the transition temperature the superconducting
gap function $\Delta\left(  \mathbf{r}\right)  $ is very small and the
\textquotedblright gap equation\textquotedblright\ can be linearized. This
linear gap equation has the form \cite{D12}, \cite{D32},\cite{L14}
\begin{equation}
\Delta\left(  \mathbf{r}\right)  =V\left(  \mathbf{r}\right)  \int
d^{3}\mathbf{r}^{\prime}\sum_{\omega_{n}}\frac{1}{\tau_{T}}H_{\omega_{n}%
}\left(  \mathbf{r,r}^{\prime}\right)  \Delta\left(  \mathbf{r}^{\prime
}\right)
\end{equation}
where $\Delta\left(  \mathbf{r}\right)  $ is the gap function at the position
$\mathbf{r}$, $\omega_{n}=\left(  2n+1\right)  \pi k_{B}T/\hbar$ are the
Matsubara frequencies, $1/\tau_{T}=2\pi k_{B}T/\hbar$, $V\left(
\mathbf{r}\right)  $ is the effective electron-electron interaction at the
position $\mathbf{r}$. The function $H_{\omega_{n}}\left(  \mathbf{r,r}%
^{\prime}\right)  $ is the product of the two single electron Green functions
$G_{\omega_{n}}\left(  \mathbf{r,r}^{\prime}\right)  $ forming a Cooperon. The
$G_{w_{n}}$ represent the amplitude of an electron traveling from
$\mathbf{r}^{\prime}$ to $\mathbf{r}$; the product describes the pair
amplitude for the Cooperon to travel from $\mathbf{r}^{\prime}$ to
$\mathbf{r}$. Since the two $G_{w_{n}}$are conjugate complex to each other,
the pair amplitude is identical to the probability for a single electron to
travel from $\mathbf{r}^{\prime}$ to $\mathbf{r}$. This equivalence of the
pair amplitude with the propagation of a single electron provides us with a
profound physical intuition for the underlying physics because it is
considerably easier to think about the propagation of a single electron than a
pair amplitude. The function $H_{\omega_{n}}\left(  \mathbf{r,r}^{\prime
}\right)  $ can be expressed by the function $F\left(  \mathbf{r,}%
0\mathbf{;r}^{\prime},t^{\prime}\right)  $ which gives the probability of a
single electron to travel from $\mathbf{r}^{\prime}$ to $\mathbf{r}$ during
the time interval $\left\vert t^{\prime}\right\vert $ (departing at
$\mathbf{r}^{\prime}$ at the negative time $t^{\prime}<0$ and arriving at
$\mathbf{r}$ at the time $t=0$) while it experiences an exponential damping of
$\exp\left(  -2\left\vert \omega_{n}\right\vert \left\vert t^{\prime
}\right\vert \right)  $.
\[
H_{\omega_{n}}\left(  \mathbf{r,r}^{\prime}\right)  =\int_{-\infty}%
^{0}dt^{\prime}e^{2\left\vert \omega_{n}\right\vert t^{\prime}}F\left(
\mathbf{r,}0\mathbf{;r}^{\prime},t^{\prime}\right)  N\left(  \mathbf{r}%
^{\prime}\right)
\]
with $N\left(  \mathbf{r}^{\prime}\right)  $ is the (BCS)-density of states
for one spin direction.
\begin{align*}
\frac{1}{\tau_{T}}\sum_{\omega_{n}}H_{\omega_{n}}\left(  \mathbf{r,r}^{\prime
}\right)   &  =\sum_{\omega_{n}}\int_{-\infty}^{0}\frac{dt^{\prime}}{\tau_{T}%
}e^{2\left\vert \omega_{n}\right\vert t^{\prime}}F\left(  \mathbf{r,}%
0\mathbf{;r}^{\prime},t^{\prime}\right)  N\left(  \mathbf{r}^{\prime}\right)
\\
\tau_{T}  &  =\frac{\hbar}{2\pi k_{B}T}\text{, }n_{c}=\frac{\Theta_{D}}{2\pi
T}%
\end{align*}
The function
\[
\sum_{\omega_{n}}e^{-2\left\vert \omega_{n}\right\vert t^{\prime}/\tau_{T}%
}=2\sum_{n=0}^{n_{c}}e^{-\left(  2n+1\right)  x}=2\frac{1-e^{-2\left(
n_{c}+1\right)  x}}{\sinh\left(  x\right)  }%
\]
describes the exponentially decaying coherence of the Cooperons. The resulting
gap equation is
\begin{equation}
\Delta\left(  \mathbf{r}\right)  =V\left(  \mathbf{r}\right)  \int
d^{3}\mathbf{r}^{\prime}\int_{-\infty}^{0}\frac{dt^{\prime}}{\tau_{T}}%
\sum_{\omega_{n}}e^{2\left\vert \omega_{n}\right\vert t^{\prime}}F\left(
\mathbf{r,0;r}^{\prime},t^{\prime}\right)  N\left(  \mathbf{r}^{\prime
}\right)  \Delta\left(  \mathbf{r}^{\prime}\right) \nonumber
\end{equation}
One can describe the physical meaning of the above equation as follows.

\begin{itemize}
\item  At the (negative) time $t^{\prime}$ we have $N\left(  \mathbf{r}%
^{\prime}\right)  \Delta\left(  \mathbf{r}^{\prime}\right)  $ electrons
departing from the position $\mathbf{r}^{\prime}$.

\item  A fraction $F\left(  \mathbf{r,}0\mathbf{;r}^{\prime},t^{\prime
}\right)  $ of them reaches the position $\mathbf{r}$ at the time $t=0$.

\item  During this propagation their number is decaying as $\sum_{\omega_{n}%
}e^{2\left|  \omega_{n}\right|  t^{\prime}}$ (due to the dephasing of the pair
amplitude). In this sum the term $\exp\left(  2\left|  \omega_{0}\right|
t^{\prime}\right)  =\exp\left(  -2\pi k_{B}T\left|  t^{\prime}\right|
/\hbar\right)  $ is the most important one because it represents the longest coherence-time

\item  At the time $t=0$ and position $\mathbf{r}$ we integrate over all
initial starting positions and over all times $t^{\prime}<0$.

\item  The resulting (number of) electrons at $\left(  \mathbf{r,}t=0\right)
$ multiplied with the attractive electron-electron interaction $V\left(
\mathbf{r}\right)  $ yields the gap function at the position $\mathbf{r}$.
\end{itemize}

(One of the authors derived the time-dependent Ginzburg-Landau equation from
this interpretation by replacing the time $0$ by $t$ and $\Delta\left(
\mathbf{r}\right)  ,\Delta\left(  \mathbf{r}^{\prime}\right)  $ by
$\Delta\left(  \mathbf{r,}t\right)  ,\Delta\left(  \mathbf{r}^{\prime
},t^{\prime}\right)  $ \cite{B22}.)

In a homogeneous superconductor one has a constant $\Delta\left(
\mathbf{r}^{\prime}\right)  $. Then the integral $\int d^{3}\mathbf{r}%
^{\prime}F\left(  \mathbf{r,0;r}^{\prime},t^{\prime}\right)  $ is identical to
one, (independent of $t^{\prime})$, the time integration over $dt^{\prime}$
yields the sum over the inverse Matsubara frequencies $\sum_{\omega_{n}}%
\int_{-\infty}^{0}dt^{\prime}e^{2\left|  \omega_{n}\right|  t^{\prime}}%
=\sum_{\omega_{n}}\frac{1}{2\left|  \omega_{n}\right|  },$ and one obtains
Gorkov's condition for the transition temperature.

For a sandwich of a superconductor and normal conductor the gap function
vanishes in the normal conductor because the effective electron-electron
interaction is zero. Therefore one has to consider only electrons which start
and arrive in the superconductor, i.e. the contribution of $F\left(
\mathbf{r,0;r}^{\prime},t^{\prime}\right)  $ only counts when $\mathbf{r}%
^{\prime}$ and $\mathbf{r}$ lie in the superconductor. This contribution
$F\left(  \mathbf{r,0;r}^{\prime},t^{\prime}\right)  $ is continuously reduced
when a normal conductor film is condensed onto the superconductor because
along the way the electrons can escape into the normal film. This reduces the
contribution of the integral over $\int d\mathbf{r}^{\prime}\int_{-\infty}%
^{0}dt^{\prime}$ and destroys the self-consistency of the gap equation. To
maintain the self-consistency one has to lower the temperature so that the
dephasing is reduced.

In Fig.5 several paths are shown. The start points of the paths represent the
points $\mathbf{r}^{\prime}$ at the time $t^{\prime}$ and the arrows the
points $\mathbf{r}$ at the time $t=0$. Only the Path $A$ contributes to the
gap equation. Its contribution increases when the thickness of the normal
conductor N is reduced and (or) the density of states in N is decreased
(because then the transmission from S to N is reduced).

$%
{\parbox[b]{2.2217in}{\begin{center}
\includegraphics[
height=2.7347in,
width=2.2217in
]%
{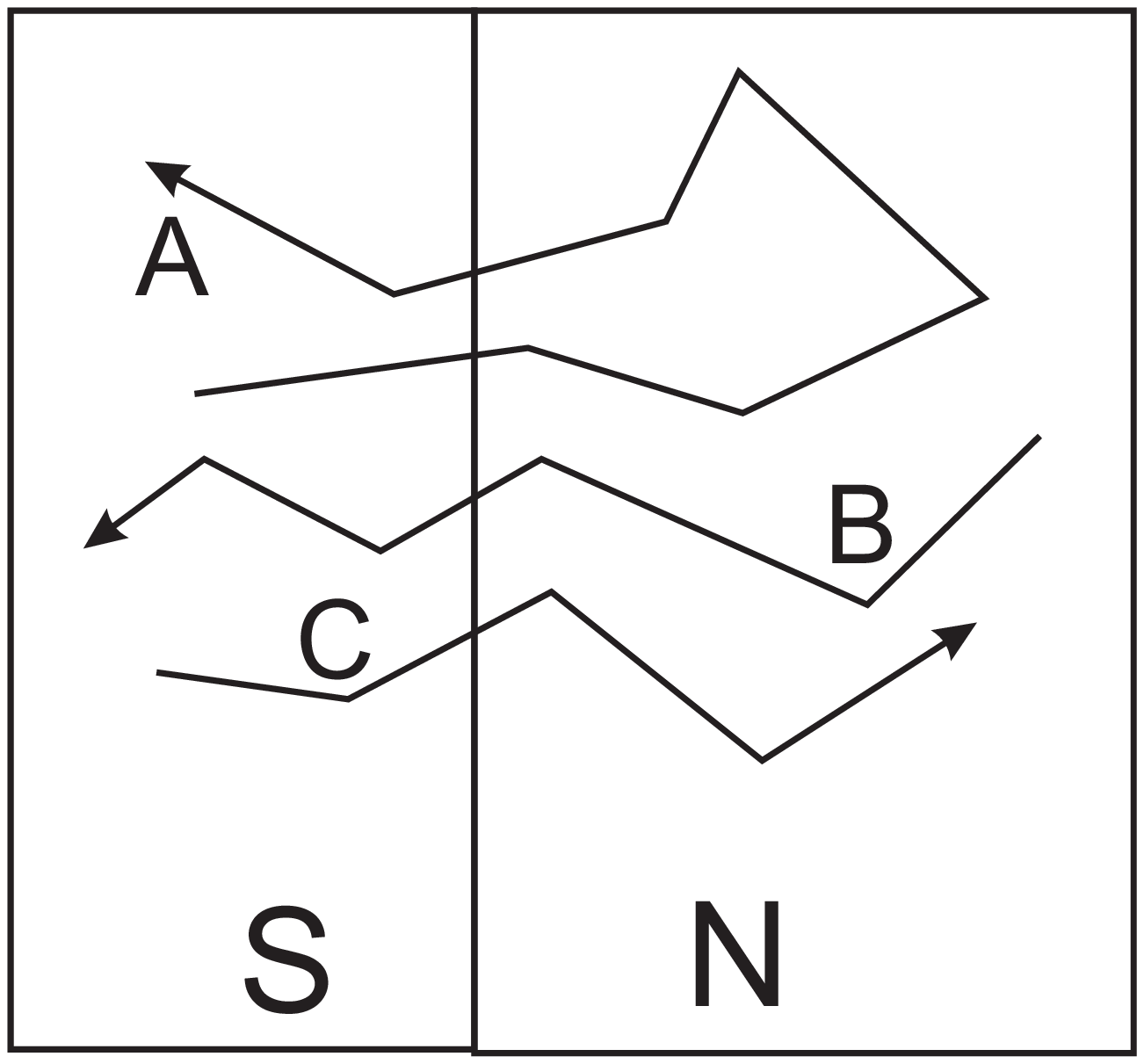}%
\\
Fig.5: The propagaton of single electrons (corresponding to the pair
amplitude) in an super- normal conductor sandwich. Only the path A contributes
to the gap function.
\end{center}}}%
$

Since $T_{c}$ does not change when sub-mono-layers of Pb are condensed on the
K film we conclude that neither the effective thickness of the K nor its
density of states is reduced. On the other hand the resistance measurements
show that there is practically no conductance through the K film covered with
some Pb. The K film behaves as if its electrons are unable to carry a current
in the x-y-plane but can easily move in the z-direction.

We like to demonstrate with an example that such a behavior is not unheard of
(without claiming that this example explains nor that it is relevant for the
present experiment). Suppose that the Pb impurities form a periodic
superlattice on the surface of the K. This will impose a two-dimensional
periodicity on the K film. Such a periodicity has been observed in the
literature \cite{C18}. It will introduce band gaps in the Fermi surface in
$k_{x}$- and $k_{y}$-direction. As a consequence the conductance in the
x-y-plane could be strongly reduced, even suppressed while the $k_{z}%
$-direction is not effected.

In our experiment localization effects are more likely. The question is if we
have localization is it

\begin{itemize}
\item  due to disorder? If the Pb atoms on the K surface are disordered and
cause Friedel oscillations in the film these might impose disorder onto the
position of the K atoms. Since the alkali metals are open metal its atoms
should be easily displaced by charge oscillations.

\item  due to the large electron-electron interaction in the alkali metals in
combination with the structural disorder?

\item  due to charge density waves \cite{O17}?
\end{itemize}

In this paper we investigated the effect of the second (upper) interface
between K and Pb. There is also some effect from the first (lower) interface
between K and Pb. Additional research will be required to better understand
this phenomena. We are presently investigating the superconducting proximity
effect between Pb and different alkali metals quantitatively.

Acknowledgment: The research was supported by NSF Grant No. DMR-0124422.

\end{document}